\documentclass[useAMS,usenatbib]{mn2e}

\usepackage{graphicx}
\usepackage{times}
\usepackage{amssymb}
\usepackage{natbib}

\usepackage[T1]{fontenc}
\usepackage{aecompl}

\newcommand{\be}{\begin{equation}}
\newcommand{\ee}{\end{equation}}

\def\ltsima{$\; \buildrel < \over \sim \;$}
\def\simlt{\lower.5ex\hbox{\ltsima}}
\def\gtsima{$\; \buildrel > \over \sim \;$}
\def\simgt{\lower.5ex\hbox{\gtsima}}

\title[Counterrotating Stars in Stellar Disks]{Counterrotating Stars in
Simulated Galaxy Disks}

\author[Algorry et al.]{
David G. ~Algorry,$^{1}$
Julio F. ~Navarro,$^{2}$
Mario G. ~Abadi,$^{1}$
Laura V. ~Sales,$^{3}$
\newauthor 
\ Matthias ~Steinmetz,$^{4}$ and
Franziska ~Piontek$^{5}$\\
$^{1}$ Observatorio Astron\'omico de C\'ordoba and CONICET, C\'ordoba, Argentina\\
$^{2}$ Senior CIfAR Felloe. University of Victoria. Victoria, BC Canada V8P5C2\\
$^{3}$ Max-Planck Institute for Astrophysics, Karl-Schwarzschild-Strasse 1, 85740 Garching, Germany\\
$^{4}$ Astrophysikalisches Institut Potsdam, An der Sternwarte 16, 14482 Potsdam, Germany\\
$^{5}$ Potsdam Institute for Climate Impact Research, Telegraphenberg A31, 14473 Potsdam \\
}

\begin{document}

\date{}

\pubyear{2013}

\label{firstpage}

\maketitle

\begin{abstract} 
  Counterrotating stars in disk galaxies are a puzzling dynamical
  feature whose origin has been ascribed to either satellite accretion
  events or to disk instabilities triggered by deviations from
  axisymmetry. We use a cosmological simulation of the formation of a
  disk galaxy to show that counterrotating stellar disk components may arise
  naturally in hierarchically-clustering scenarios even in the absence
  of merging. The simulated disk galaxy consists of two coplanar,
  overlapping stellar components with opposite spins: an inner
  counterrotating bar-like structure made up mostly of old stars
  surrounded by an extended, rotationally-supported disk of younger
  stars.  The opposite-spin components originate from material
  accreted from two distinct filamentary structures which at turn
  around, when their net spin is acquired, intersect delineating a ``V''-like
  structure. Each filament torques the other in opposite
  directions; the filament that first drains into the galaxy forms the
  inner counterrotating bar, while material accreted from the other
  filament forms the outer disk. Mergers do not play a substantial
  role and most stars in the galaxy are formed {\it in situ}; only
  $9\%$ of all stars are contributed by accretion events.  The
  formation scenario we describe here implies a significant age
  difference between the co- and counterrotating components, which
  may be used to discriminate between competing scenarios for the
  origin of counterrotating stars in disk galaxies.
 \end{abstract}

\section{Introduction}
\label{sec:intro}

%%%%%%%%%%%%%%%%%%%%%%%%%%%%%%%%%%%%%%%%%%%
\begin{figure*}
\begin{center}
\includegraphics[width=\linewidth,clip]{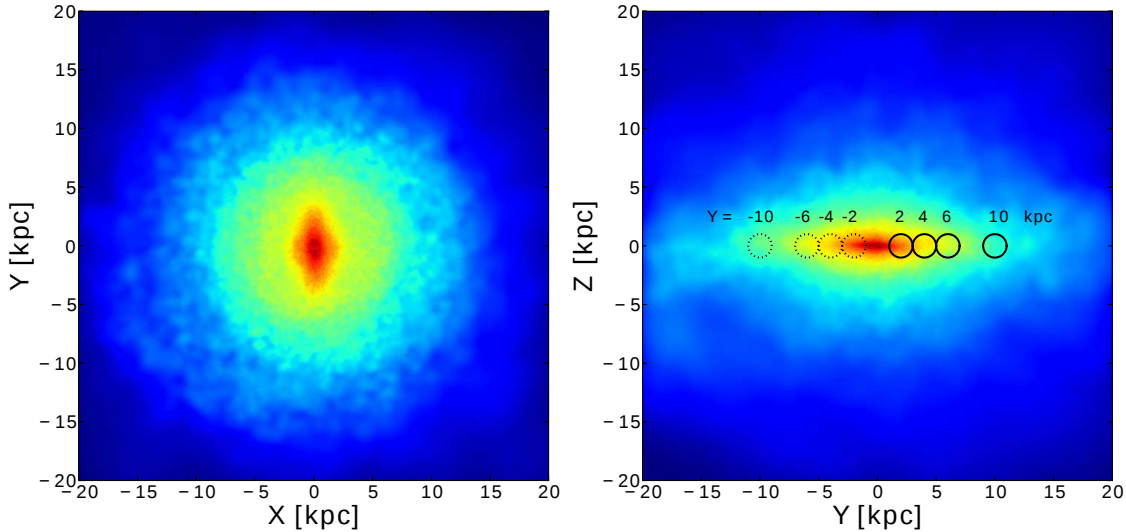}
\end{center}
\caption{Face-on (left) and edge-on (rightl) stellar 
surface density maps of the simulated galaxy at 
 $z=0$. Note the inner bar-like structure clearly seen in the face-on
 view. The edge-on view shows the bar on its side. The small
 circles along the major axis indicate the lines of sight
 corresponding to the velocity distributions shown in Fig.~\ref{Fig2LOSVD}.
\label{FigProjDensMaps}}
\end{figure*} 
%%%%%%%%%%%%%%%%%%%%%%%%%%%%%%%%%%%%%%%%%%%

Galaxy disks are rotationally-supported systems that consist largely
of stars in co-planar, nearly-circular orbits. Although most disks
also contain stars in counterrotating orbits, these typically belong
to separate spheroidal components (a bulge or a halo) characterized by
little or no net rotation and large orbital excursions from the disk
plane. Stellar disks are the unmistakable result of dissipative
collapse, where a gaseous component cools and settles onto a
rotationally-supported structure before turning gradually into
stars. The spheroidal components, on the other hand, are thought to
arise from merger events where existing stars have their orbits
randomized by the fluctuating gravitational potential.

Infrequently, a population of stars with small vertical motions but
that counterrotates relative to the prevailing spin is also detected
through the presence of separate peaks in the line-of-sight velocity
distribution (LOSVD) measured from high-resolution
spectra. Counterrotating stars are sometimes confined to the inner
regions of the galaxy, as is the case for the counterrotating bulge of
NGC 7331 \citep{Prada1996}, but are also found on occasion to overlap
the main disk component, as in NGC 5719
\citep[]{Vergani2007,Coccato2011}; NGC 7217 \citep{Merrifield1994};
NGC 3593 \citep{Bertola1996}; and NGC 4550 \citep{Rubin1992,Rix1992}.

The origin of such populations is not well understood. An external
origin has been predicated on the basis that counterrotating bulges
are reminiscent of the dynamically-decoupled cores present in some
elliptical galaxies \citep{Bender1992,Carter1998,Emsellem2007,Kuntschner2010},
which may originate from the accretion
of compact satellites able to survive disruption and that settle at the
center of the main galaxy (\citealp{Balcells1990}; but see
\citealt{VandenBosch2008}). In the case of a disk galaxy, dynamical
friction can drag a satellite to the disk plane and circularize its
orbit before disruption, leaving behind a population of co- or
counterrotating disk stars, depending on the initial orbital angular
momentum of the satellite \citep{Quinn1993,Abadi2003b}.

An alternative scenario envisions counterrotating stars as a result of 
the accretion of mainly gaseous material with angular momentum opposite
that of the pre-existing stellar disk. Given its collisional nature, gas would
quickly settle onto a counterrotating disk that would gradually turn
into stars. The final result would be the overlap of two stellar disks
sharing the same orbital plane.

How can we discriminate between these two scenarios? The gas-accretion
scenario makes specific predictions about the angular momentum,
spatial distribution, and age difference of the two disks. Given the
collisional nature of accreted gas, the two structures cannot form
simultaneously, implying a well-defined age difference. Besides, since
angular momentum typically correlate with accretion time
\citep[e.g.,][]{Navarro1997b}, one would generally expect the younger
disk (accreted later) to have higher angular momentum and thus be
significantly more spatially extended than the older one. Finally, it
would be difficult for the pre-existing disk to avoid instabilities
triggered by the counter-streaming material, which may lead to the
formation of a counterrotating central bar
\citep{Palmer2003,Sellwood1994}.

We explore these issues here using a cosmological simulation of the
formation of a disk galaxy in the current paradigm of structure
formation, the $\Lambda$CDM scenario. The LOSVD of the simulated
galaxy at $z=0$ shows two separate peaks at positive and
negative velocities, a clear signature of the presence of two disk
components with opposite spins. We study their properties at the present
day, and trace them back to the moment of maximum expansion (the
turnaround, when their angular momenta are acquired) in order to
identify the origin of this unusual configuration.  We describe the
numerical simulation in Sec.~\ref{SecNumExp} and analyze its main
results in Sec.~\ref{SecRes}.  Sec.~\ref{SecConc} summarizes our main
conclusions.

\section{The Numerical Simulation}
\label{SecNumExp}

The simulated galaxy analyzed in this work was presented in detail by
\citet{Piontek2011} where we refer the interested reader for further
technical details.  In brief, the simulation zooms-in on a region
destined to form a galaxy-sized halo, identified in a $\Lambda$CDM
simulation of a large cosmological box ($64$ Mpc/$h$ on a side). The
original simulation assumes the following cosmological parameters:
H$_{0}$=73 km s$^{-1}$ Mpc$^{-1}$ (i.e., $h=0.73$), $\sigma_{8}= 0.75$, $n=0.9$,
$\Omega_{0}$=0.24 and $\Omega_{\Lambda}$=0.76, adjusted to fit the
WMAP3 results \citep{Spergel2007}.  
Both simulations were performed with the GADGET2 code
\citep{Springel2005b}. The zoomed-in simulation splits the matter
content into dark matter ($\Omega_{\rm CDM}=0.20$) and baryons
($\Omega_b=0.04$), which are initially represented by gas particle that
can be later converted into stars. GADGET implements the Smoothed
Particle Hydrodynamics (SPH) technique \citep[]{Gingold1977, Lucy1977}
to follow the hydrodynamical evolution of the gas, including pressure
gradients, shocks, and radiative cooling as described by
\citet{Katz1996}. The star formation model follows \citet{Katz1992}
 and \citet{Steinmetz1994,Steinmetz1995},
adopting a star formation rate regulated by the \cite{Schmidt1959}
law.

The gas and dark matter particle masses of the resimulation are $4.85
\times 10^5 M_{\odot}$ and $2.30 \times 10^6 M_{\odot}$,
respectively. The gravitational softening is $1.03 $ kpc for
gas, and $1.37 $ kpc for dark matter particles. Star particles 
are created with a mass $2.42 \times 10^5 M_{\odot}$,
which corresponds to half that of a gas particle. 

At $z=0$ the simulated galaxy resides in a halo with a virial radius
of $r_{200}$= $212 $ kpc (defined as the radius where the mean density
contrast $\rho(<r_{200})/\rho_{crit}$= 200), corresponding to a virial
mass $M_{200}=5.50 \times 10^{11} M_{\odot}$ and to a virial velocity
$V_{200}=105$ km s$^{-1}$.  The dimensionless spin parameter of the
halo is quite low ($\lambda \sim 0.02$) and its assembly history is
rather calm; its last major merger (defined by a mass ratio of 1:3 and
larger) happened at a relatively high redshift, $z \sim 6$.  This run
corresponds to the halo identified as {\it DM\_hr4} of the {\it
  standard} implementation of the feedback in \citet{Piontek2011} (see
their Tables 1 and 3).

Within the virial radius at $z=0$ there are 160,644, 46,067 and
214,059 star, gas and dark matter particles, respectively. Most
stars are confined within a smaller radius near the center of the
halo; we therefore define a ``galaxy radius'' , $r_{\rm gal}=20$ kpc, which
contains 150,733, 11,168 and 40,833 star, gas and dark matter
particles, respectively.

%%%%%%%%%%%%%%%%%%%%%%%%%%%%%%%%%%%%%%%%%%%%%%%%%%%%%%%%%%%%%%%
\begin{figure}
\begin{center}
\includegraphics[width=\linewidth,clip]{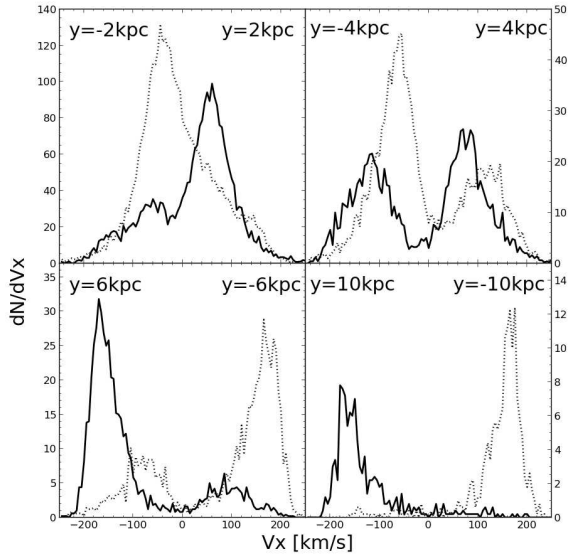}
\end{center}
\caption{Velocity distributions of stars along the lines of sight
  indicated in the edge-on projection of the galaxy in
  Fig.~\ref{FigProjDensMaps}. Solid and dotted curves correspond to lines
  of sight to the right and to the left of the center of the galaxy,
  respectively. Each panel is labeled by the distance to the center
  from each line of sight. Note the double peaks in the velocity
  distributions of the inner regions, where the lines of sight
  intersect the counterrotating bar-like central feature.
  \label{Fig2LOSVD}}
\end{figure} 
%%%%%%%%%%%%%%%%%%%%%%%%%%%%%%%%%%%%%%%%%%%%%%%%%%%%%%%%%%%%%%%

\section{Analysis and Results}
\label{SecRes}

Fig.~\ref{FigProjDensMaps} shows the surface density of stars of the simulated
disk galaxy at $z=0$, projected ``face-on'' (left panel) and ``edge-on''
(right) according to the total angular momentum of stars within
$r_{\rm gal}$, which we use to define the $Z$ axis perpendicular to the disk
plane. The face-on view shows clearly the presence of a central bar,
which we choose to align the $Y$ axis of the projection.

The bar counterrotates relative to the outer disk, as shown by the
line-of-sight velocity distribution (LOSVD) of stars in the edge-on
projection. Each velocity histogram in Fig.~\ref{Fig2LOSVD}
corresponds to stars projected within the small circles indicated in
the right-hand panel of Fig.~\ref{FigProjDensMaps}. These velocity
distributions show that the inner and outer galaxy rotate in opposite
directions, and that at intermediate radii the co- and
counterrotating components overlap. This is reminiscent of the LOSVD
of galaxies with observed counterrotating disks \citep[see,
e.g.,][]{Rix1992,Prada1996,Prada1999}.

%%%%%%%%%%%%%%%%%%%%%%%%%%%%%%%%%%%%%
\begin{figure}
\begin{center}
\includegraphics[width=1.0\linewidth,clip]{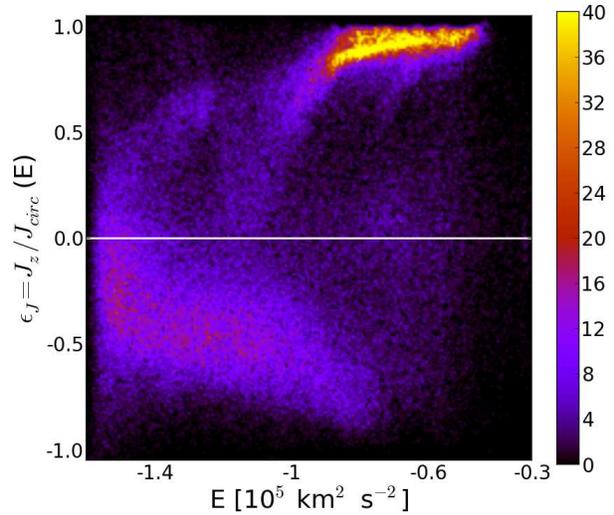}
\end{center}
\caption{ Circularity, $\epsilon_J=J_z/J_{\rm circ}(E)$, versus
  specific binding energy, $E$, of stars within $20$ kpc from the center of the galaxy. 
  Colors indicate the number of stars in each  pixel, following the
  scale on the color bar on the right. The  disk dominates in the outer
  regions and consists largely of particles on nearly-circular
  orbits. The counterrotating bar-like feature dominates in the inner
  regions. The two components show substantial overlap in binding
  energy/radius.
  \label{FigJzvsE}}
\end{figure} 
%%%%%%%%%%%%%%%%%%%%%%%%%%%%%%%%%%%%

The two components separate neatly in Fig.~\ref{FigJzvsE}, where we
show the circularity parameter, $\epsilon_J=J_{z}/J_{circ} (E)$, of
stars within $r_{\rm gal}$ as a function of binding energy,
$E$\footnote{Binding energies are computed using all particles within
  the virial radius of the halo.}. The circularity is defined as the
ratio between the $Z$-component of the specific angular momentum of a
star and that of a circular orbit with the same binding
energy \citep{Abadi2003b}. It approaches unity for co-rotating
circular orbits and $-1$ for counterrotating ones.

%%%%%%%%%%%%%%%%%%%%%%%%%%%%%%%%%
Fig.~\ref{FigJzvsE} shows that the outer disk has circularity
approaching unity and that the bar is made up predominantly of
highly-bound, counterrotating ($\epsilon_J\sim -0.5$) stars. We shall
hereafter consider  all stars with positive or negative circularity as
part of the co- or counterrotating disk, respectively. 
Each of these two components contributes about half of the total stellar
mass within $r_{\rm gal}$.
Counterrotating stars have lower energies and are more centrally
concentrated; its half-mass radius is $2.4$ kpc, compared to $5.9$ kpc
for the co-rotating disk. 
%%%%%%%%%%%%%%%%%%%%%%%%%%%%%%%%%%%%%%%%%%%%%%%%%%

\begin{figure}
\begin{center}
\includegraphics[width=0.85\linewidth,clip]{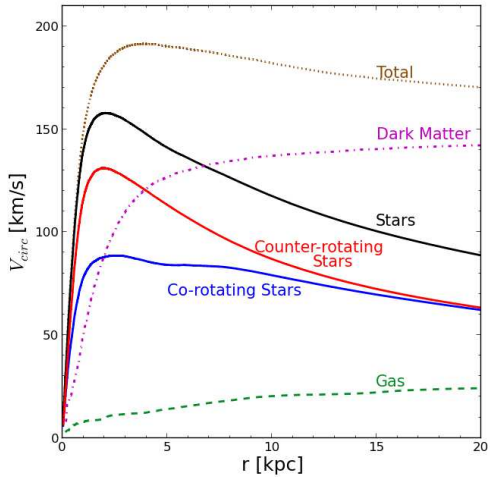}
\end{center}
\caption{Contribution to the circular
  velocity profile of the different components. Note that stars
  dominate in the inner $\sim 7$ kpc and that
  the co- and counterrotating stellar components have approximately
  the same  mass. The circular velocity profile is nearly flat, with
  the dark matter dominating the outer regions. The gas makes up less
  than $10\%$ of the total baryonic mass within $20$ kpc.
  \label{FigVcProf}}
\end{figure}

Fig.~\ref{FigVcProf}, where we plot the contribution of the various
components to the circular velocity profile of the galaxy, indicate
that stars dominate the gravitational potential within $\sim 5$
kpc. The remaining gas makes up only $10\%$ of the total baryonic
matter within $r_{\rm gal}$, has a half-mass radius of $13.8$ kpc, and
it all co-rotates with the outer disk.

\begin{figure}
\begin{center}
\includegraphics[width=1.\linewidth,clip]{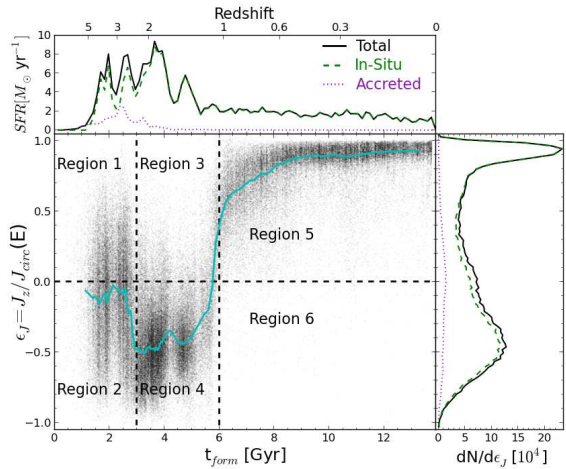}
\end{center}
\caption{{\it Main panel:} Circularity ($\epsilon_J= J_z/J_{circ}(E)$) as
  function of formation time, $t_{\rm form}$, for stars within $20$ kpc from the center
  of the galaxy at $z=0$. The cyan curve indicates the median
  $\epsilon_J$ as a function of  $t_{\rm form}$. Note the sharp age
  difference between the co- and counterrotating stellar
  components. {\it Top panel:} Distribution of star formation times,
  split by stars formed {\it in situ} (i.e., in the main progenitor)
  and those accreted from satellites. The vast majority of stars
  ($91\%$) in the galaxy were formed {\it in situ}. {\it Right panel:}
  Distribution of circularities of stars shown in the main panel.
\label{FigJzTf}}
\end{figure}

%{\bf 
The main panel of Fig. \ref{FigJzTf} shows the dependence of the
circularity parameter on the formation time of a star, and makes clear
that the co- and counterrotating components have different ages. As
expected, the younger component ($t_{\rm form}>6$ Gyr) has higher
angular momentum and populates the outer, co-rotating disk, whereas
most counterrotating (bar) stars form early ($t_{\rm form}<6$ Gyr).

\begin{figure*}
\begin{center}
\includegraphics[width=1.\linewidth,clip]{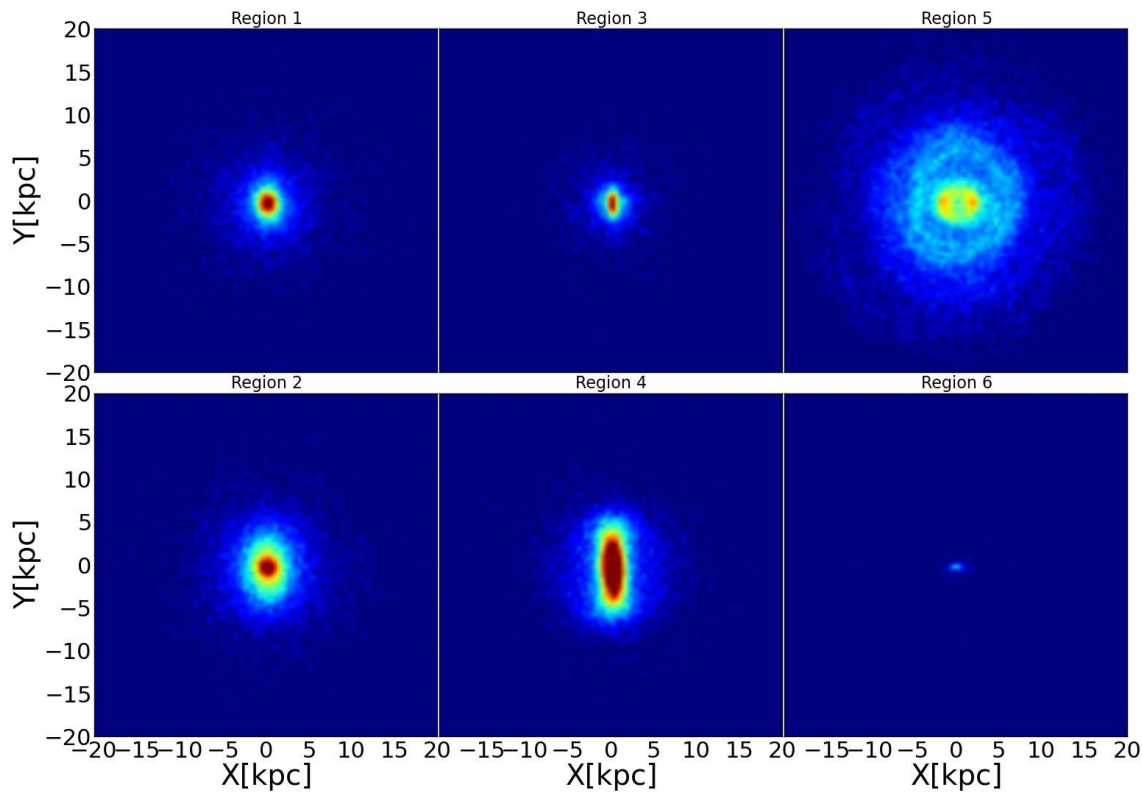}
\end{center}
\caption{As Fig.~\ref{FigProjDensMaps}, but for stars in the six regions defined 
in Fig.~\ref{FigJzTf}. Note that the counterrotating bar
has many particles with positive circularities (Region 3). Note as
well the dynamical features (rings, overdensities) in the co-rotating disk
caused by the counterrotating bar potential. 
\label{figxy}}
\end{figure*}

The presence of the bar (which grows gradually after $t=6$ Gyr and is
fully formed by $t=9$ Gyr) has interesting effects on the orbits of
co- and counterrotating stars. We study this by dividing them into
three subgroups each according to their age: $t_{\rm form}<3$ Gyr, 3
Gyr $<t_{\rm form}<$6 Gyr, and $t_{\rm form}>6$ Gyr (see the six
regions outlined in Fig.~\ref{FigJzTf}).  Fig. \ref{figxy} shows the
spatial distribution of each of these subcomponents as seen in a
face-on view of the galaxy. The oldest two (regions 1 and 2) form a
central ``bulge'' that is affected little by the bar and that contains
most of the accreted stars.

Fig.~\ref{figxy} also shows that even co-rotating stars contribute to
the bar pattern; indeed, stars in region 3 follow closely the bar
pattern even though they rotate in {\it opposite} sense to that of
most stars in the bar, as predicted by theory
\citep[e.g.,][]{Zhao1996,Wozniak1997}. Young stars (regions 5 and 6)
define the main sense of rotation of the galaxy, but their spatial
distribution reveals that their orbits are significantly affected by
the bar. Note, for example, the ``ring'' feature at $R\sim 6$ kpc
visible in region 5, and the presence of two overdense regions of
stars along a direction perpendicular to the bar (at $X\sim \pm 2$
kpc). These patterns likely correspond to the location of resonances
in the prograde disk with the counterrotating bar potential,
highlighting a complex dynamical situation which we plan to study in
detail in a future contribution.  We turn our attention now to the
origin of the co- and counterrotating disks.

As discussed above, counterrotating stars are mainly $\sim 6$ Gyr old
or older, and some of them have been accreted from different
progenitors. The latter population, however, make up only a small
fraction, $9\%$, of the stellar mass of the galaxy. Indeed, both the
co- and counterrotating disks are made up primarily of stars formed
{\it in-situ} and, given their spatial overlap, must differ in their
formation time.  This suggests that the origin of these two components
is linked to differences in the angular momentum of the accreting gas
which, at late times, flows with a net spin opposite to that of the
preexisting galaxy.
%}

%%%%%%%%%%%%%%%%%%%%%%%%%%%%%%%%%%%%%
%%%%%%%%%%%%%%%%%%%%%%%%%%%%%%%%%%%%%

%%%%%%%%%%%%%%%%%%%%%%%%%%%%%%%%%%%%%%%%%%%%%%%%%%%%%%%

What causes the change in spin direction of the accreting gas? We look
for clues in the temporal evolution of the two components, going back
to the time of maximum expansion (i.e., turnaround), when their net
angular momentum was acquired. To this aim, we computed the evolution
of the angular momentum of the baryons making up each of the two
components as a function of time, in the reference frame of the main
galaxy progenitor. As reported in earlier work \citep[see,
e.g.,][]{Navarro1997b, Abadi2010}, the angular momentum of each
component rises quickly with time until turnaround and stays roughly
constant or declines slightly thereafter.

\begin{figure*}
\begin{center}
\includegraphics[width=0.92\linewidth,clip]{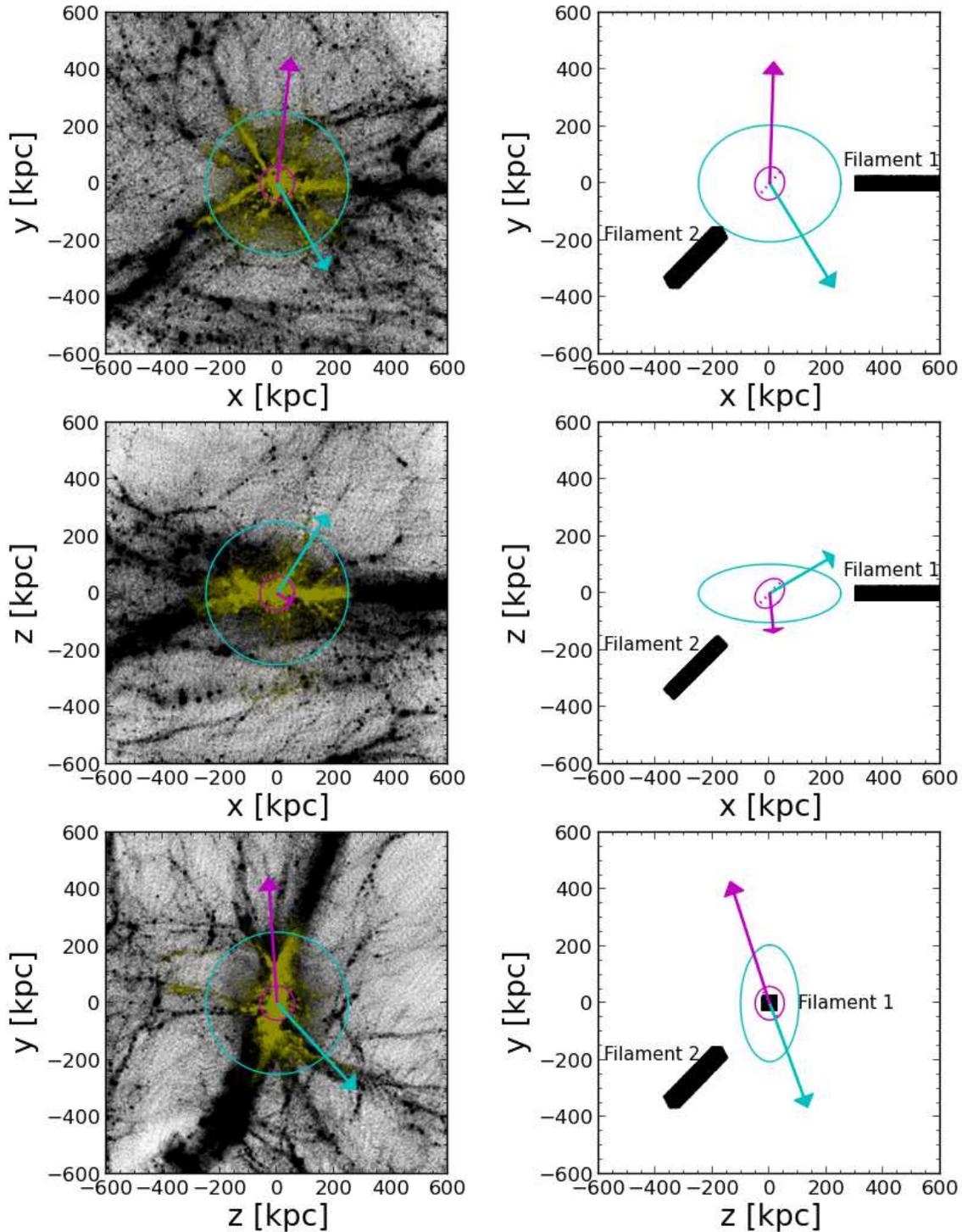}
\end{center}
\caption{{\it Left-hand panels:} Spatial distribution of all the
  mass (black dots) and the baryons destined to form the galaxy
  at z=0 (yellow dots) at $z_{\rm ta}=2.12$  inside a box of $1.2$ Mpc (physical).
  The cyan and magenta circles contain 90\% and 45\% of all yellow
  particles, respectively.  Material within the outer/inner circle
  contributes mostly to the co- and counterrrotating disk at $z=0$. Arrows show the orientation of the torque
  vector due to the external material operating on the baryonic
  particles inside the circles of each respective color, as measured
  in the simulation. Note that the torque on the inner/outer galaxy
  points in nearly opposite directions. {\it
    Right-hand panels:} Toy model of the mass distribution shown in
  the left-hand panels. The filaments represent the particles
  which generate the external tides. Magenta and cyan ellipsoids
  indicate the shape and orientation of the material that will form
  the inner and outer galaxy, respectively. Arrows indicate the tidal
  torque vector exerted by the external mass distribution, as measured
  in the model. Note the excellent agreement between the orientation of the tidal torque
  vector in the right- and left-hand panels.
  \label{fig7}}
\end{figure*}

Interestingly, at turnaround ($z_{\rm ta}=2.12$) the two 
components already had opposite spins, implying that something in the
torquing process that endows each component with net rotation is
responsible for the origin of the two components. We show this in
Fig.~\ref{fig7}, where we plot the spatial configuration of all
particles in a $1.2$ Mpc (physical) box centered on the galaxy main
progenitor at $z_{\rm ta}$. We highlight in yellow the baryons that at
$z=0$ will be found within the galaxy radius, $r_{\rm gal}$, and plot
them in a Cartesian ($x$,$y$,$z$) reference frame aligned with the
principal axes of their inertia tensor ($x$ and $z$ correspond to the
major and minor axis, respectively).  Roughly $90\%$ of all baryons
destined to form the galaxy are contained at $z_{\rm ta}=2.12$ within the cyan
circle shown in the left-hand panels of Fig.~\ref{fig7}, and about
half of them are found within the magenta circle. With few exceptions,
baryons within the inner magenta circle contribute to the
counterrotating component at $z=0$; those outside it make up
primarily the outer disk. The inner baryons collapse to form a 
stellar disk which, after the accretion of the outer component, 
turns into the counterrotating bar seen at $z=0$.

%%%%%%%%%%%%%%%%%%%%%%%%%%%%%%%%%%%%%

Arrows in the left panels of Fig.~\ref{fig7} indicate the direction of
the torque operated on each of the inner (magenta) or outer (cyan)
baryons by external material (i.e., that outside the cyan
circle). They clearly point in opposite directions ($\sim
130^{\circ}$), mainly along the $y$ (intermediate) axis of the
projection, as expected from tidal torque theory. Indeed, the
component of the torque along one axis, $\tau_i \approx T_{jk} (I_{jj}
- I_{kk})$ ($i\ne j\ne k$ and run from 1 to 3), scales to first order
as the difference of the inertia moments of the other two axes:
$\tau_i$ is therefore generally largest along the intermediate
axis. (Here $T_{ij}=\partial^2\phi/\partial x_i \partial x_j$ is the
tidal tensor generated by external material. The calculation assumes a
reference frame where the inertia tensor is diagonal; see
\citealt{Navarro2004b} for details.)

The reason for the switch in the direction of the torque operating on
the inner/outer galaxy becomes clear once we realize that the accreted
gas is channeled towards the galaxy primarily along two large-scale
filaments: one, which we label ``Filament 1'', is aligned with the
positive $x$ axis of Fig.~\ref{fig7} and a second one, which we call
``Filament 2'', traces the diagonal of the ($-x$,$-y$,$-z$)
octant. The galaxy forms roughly at the ``V''-like intersection of these two
filaments: Filament 1 contributes most of the material of the outer
galaxy, while the baryons making up the counterrotating inner galaxy
flow mainly along Filament 2.

The right-hand panels of Fig.~\ref{fig7} present a simplified model
that attemps to mimic the external mass distribution responsible for
torquing the material destined to be accreted into the galaxy. We
represent the inner (magenta) and outer (cyan) galaxy material at
turnaround as triaxial ellipsoids whose major axis is aligned with
that of each filament. The axial ratios of the ellipsoids are
$(b/a=0.93,c/a=0.70)$ and $(b/a=0.82,c/a=0.41)$ for the inner and
outer galaxy material, respectively. These values, together with their
orientations, are adopted to match closely those of the inertia tensor
of the inner and outer material at that time.

The filaments themselves are represented in the model by $10^{11}
M_{\odot}$ one-dimensional mass distributions that extend out to $600$
kpc from the center, and are aligned as indicated in the right-hand
panels of Fig.~\ref{fig7}. Because the principal axes of the inner
galaxy material are aligned with Filament 2, it can only be torqued by
Filament 1. Analogously, the outer galaxy material can only be torqued
by Filament 2. This arrangement produces torques of roughly opposite
signs on the inner and outer galaxy, which are shown by the arrows in
the right-hand panels of Fig.~\ref{fig7}, and
results in perfectly antialigned spins at $z=0$. Note that the directions and
relative magnitudes of the components of the torque in the model
(right-hand panels) are in very good agreement with those actually
measured in the simulation for the inner and outer galaxy (left-hand panels).

We conclude that the counterrotating components originate from
the peculiar torquing process that arises from the accretion of gas
along two different filaments that intersect in a ``V''-like
configuration at the time of turnaround. The material that flows into
the galaxy along one filament is naturally torqued in a direction
opposite to that of material accreting along the other. This would not
necessarily result in a two-component disk; if accretion along both
filaments was coeval then the collisional nature of the gas would
ensure that only one disk forms. In the case of our simulated galaxy,
however, enough time separates the two episodes of accretion to allow
the stellar component of the inner galaxy to form before the accretion
of the material that forms the outer disk occurs.

%%%%%%%%%%%%%%%%%%%%%%%%%%%%%%%%%%%%%%%%%%%%%%%%%%%%%%%%%%%%%%%%%%%%%%%%%%%%%%%%%%%%%
%%%%%%%%%%%%%%%%%%%%%%%%%%%%%%%%%%%%

\section{Summary and Conclusions}
\label{SecConc}

We explore the origin of counterrotating stars in disk galaxies whose
orbits are confined to the plane of the disk. This is a puzzling
dynamical feature known to be present in a number of disk galaxies,
but its origin, be it mergers, secular evolution, or external gas
accretion, is still under debate. We show here that this configuration
arises without fine tuning in a simulation of galaxy formation in a
$\Lambda$CDM universe and may therefore offer interesting insight into
its origin. Most stars in the simulation are formed {\it in-situ}, and
mergers play no significant role in the formation of the disk.

The co- and counterrotating components have distinct ages and, despite
some overlap, show differences in their spatial distributions. We show
that they actually arise from separate episodes of infall of gas with
opposite spins. The spin reversal may be traced to the fact that the
galaxy forms at the end of two intersecting filaments which, at
turnaround (when their net angular momentum is acquired), form a
``V''-like configuration. One filament provides the early-collapsing
material that forms the stars of the inner galaxy while the outer
galaxy material is supplied by the other. Each filament torques
material on the other filament but not on its own, leading to the
acquisition of opposite spins before collapse. The early collapsing
material is able to form an inner stellar disk before the second
episode of accretion results in the formation of a more extended disk
that rotates in the opposite direction.

Since filamentary gas accretion is commonplace in $\Lambda$CDM, one
may wonder why counterrotating stars are not more prevalent in disk
galaxies than observed. One reason is that a number of conditions are
required to form stellar disks with counterrotating
components. Although accretion along filaments is quite common in
$\Lambda$CDM, the ``V''-like configuration at turnaround that leads to
opposite spins is much less frequent. In addition, the accretion from
one and the other filament must be timed to allow for the two stellar
components to form one before the other is accreted, again a condition
that would be satisfied only in a minority of cases.

We conclude that the accretion of gas with spin misaligned relative to
the preexisting galaxy might be responsible for at least some of the
counterrotating stars in disk galaxies, especially those where such
stars differ in age and spatial distribution from the main disk. These
spin misalignments arise naturally in a hierarchically-clustering
scenario, where net angular momentum results from the coupling between
the tidal stress tensor and the inertia tensor of the material
destined to form a galaxy at early times \citep[see,
e.g.,][]{Navarro2004}. Differences in the spatial distribution of
early- and late-accreting material at early times can therefore result
in large differences in the direction of their acquired spin.  

Such misalignments may play a more significant role in galaxy
formation than is usually recognized. Indeed, they may not only
explain morphological and dynamical oddities such as counterrotating
disk stars and polar rings \citep[e.g.,][]{Maccio2006} but may also
play a substantial role in the formation of spheroidal
galaxies. Indeed, \citet{Scannapieco2009} and \citet{Sales2012} have argued that many stellar
spheroids may result not from mergers but rather from the overlap of
several episodes of gaseous accretion with misaligned spins. Although
each episode leaves behind a population of stars with a well-defined
age and spin the ensemble may be indistinguishable from a slowly
rotating elliptical galaxy. Further work should help to clarify
whether the vast diversity of galaxy morphologies is indeed linked to
the complex patterns of gas accretion in a hierarchically-clustering
universe, as these simulations suggest.

\section{ACKNOWLEDGEMENTS}
\label{anknowl}

We thank the referee, Prof. Herv\'e Wozniak, for a prompt and very constructive report.
We thank Alejandro Benitez Llambay for his code py-SPHViewer.

\bibliography{master}

\end{document}